# SOFTWARE DEVELOPMENT STANDARD AND SOFTWARE ENGINEERING PRACTICE: A CASE STUDY OF BANGLADESH

Zerina Begum, Mohammed Shafiul Alam Khan, M. Z. Hafiz, Md. Saiful Islam and Md. Shoyaib

Institute of Information Technology, University of Dhaka, Dhaka-1000, Bangladesh

## ABSTRACT

Improving software process to achieve high quality in a software development organization is the key factor to success. Bangladeshi software firms have not experienced much in this particular area in comparison to other countries. The ISO 9001 and CMM standard has become a basic part of software development. The main objectives of our study are: 1) To understand the software development process uses by the software developer firms in Bangladesh 2) To identify the development practices based on established quality standard and 3) To establish a standardized and coherent process for the development of software for a specific project. It is revealed from this research that software industries of Bangladesh are lacking in target set for software process and improvement, involvement of quality control activities, and standardize business expertise practice. This paper investigates the Bangladeshi software industry in the light of the above challenges.

## INTRODUCTION

As computer technology offers efficient and high performance information processing, it has got popularity over the home and office users in the whole world. By the decade of 1990, in Bangladesh, it has also taken an important role. Since during this time PCs become more user friendly and attractive, the number of users had been increased. Beside the general users, in Bangladesh, a number of software developers have been increased as well. Many of Computer Science and Engineering graduates form public and private universities as well as computer diplomas from training institutions are getting employed to the local software companies. As the time goes, the overall development of skill of software developers has been increased with respect to Bangladesh.

Bangladesh stands out distinctly as a potential software-exporting nation, considering the analytical and technological ability of its people. Bangladesh is one of the potential countries where software development is to be grown as a software industry. According to Bangladesh Association of Software and Information Service (BASIS), there are around 250 companies, are working closely with the development of software for local and international market for different information and communication technology services [1]. Bangladesh is a country, where the only surplus property is the human resource. Considering the earning of foreign exchanges and removing of unemployment problem, software industry is a very prospective field. To make this field more profitable, several plans has been done by the government and private organizations in the last several years. The Government of Bangladesh made an in depth study on how the software sector of the country could be designed to suit the needs of the global market. To follow up on the outcome of the study and to monitor the issues associated with the sector's growth and development, a high powered National Standing Committee (NEC) on software export has been formed. This standing committee has brought together the concerned government offices organizations and leaders of the software trade to work in unison to study the problems and prospects of the sector [2].

Table 1 shows the business application nature of software service of software industry at Bangladesh [1]. It is notable that each software company in software industry develops multiple category of software service.

**Table 1. Products/Service Range of Local Software Industry**

| Products/Services Category | % of Companies Offering Services |
|---|---|
| Accounting & Financial Management | 69% |
| Inventory Management | 59% |
| Human Resource Software | 58% |
| Web Site/Web Application Development | 57% |
| ERP (Enterprise Resource Planning) | 48% |
| Software Implementation & Integration | 46% |
| Billing | 43% |
| Asset Management | 38% |
| POS (Point of Sales) | 37% |
| E-Commerce | 36% |
| Data Entry/Data Conversion | 34% |
| CRM (Customer Relationship Management) | 32% |
| E-Governance Application | 29% |
| SCM (Supply Chain Management) | 27% |
| Data Warehousing | 23% |
| Access Control | 22% |
| Mobile/Wireless Application Development | 18% |
| E-Learning | 17% |
| Data Security | 14% |
| Gaming Software | 6% |

## METHODOLOGY OF THE STUDY

Methodology of the study discusses about the way it has been approached, how data have been collected and then analyzed. This is an exploratory study trying to find out the current software development standard of Bangladesh. Secondary studies are conducted to acquire the preliminary knowledge to explain the primary data. Data have been collected through questionnaire survey and interviews. A structured questionnaire was developed to collect relative information. There are 220 BASIS listed software firms who are directly or indirectly developing software for the local market as well as international market. To find out the existing process in place in different companies, we selected only those companies who have been doing only software development i.e. software development and sell are their main business. Convenience sampling method has been employed to select 50 software companies from Dhaka City. Chief Executive Officer (CEO) of each selected company has been asked to fill out the questionnaires. The questionnaire, which is mentioned before to be used for data collection of this study, will be distributed among the CEO through the Post Graduate Diploma in Information Technology (PGDIT), students of session 2008-2009, Institute of Information Technology, University of Dhaka, who were selected and trained them up for conducting the interview through the structured questionnaires. All the relevant information was collected during May 2008 to July 2008. At first, descriptive statistics (frequency & percentage) were presented to show the overall condition of the sample. For the above analyses, Statistical Package for Social Science (SPSS), Version 10.0.1 and MS Excel were used. Therefore, all results of this paper are developed from the primary source if not otherwise mentioned.

**SOFTWARE DEVELOPMENT PROCESS**

A software process is a framework for the tasks that are required to build high quality software [3]. Therefore, software process defines the approach that is taken as software is engineered. It may be an ad-hoc process devised by the team for one project. But the team often refers to a standardized documented methodology which has been used before on similar projects or one which is used habitually within an organization. Some managers who are held accountable for software development may seek to find the commonalities in the efforts of their organizations. If those managers are process oriented then they may seek methodologies or other proxies which can serve as templates for the software development process. Another reason why software development process is important is that a process provides organizational stability and more control to its activity.

Scott et al. [4] discusses important requirements for software process improvement (SPI) that should be built on developed and proven SPI technologies like CMM, SPICE, and ISO 12207. However, as these models are built actually for large companies, they may not be appropriate for Small and Medium Entrepreneurs (SMEs), where they have short deadlines, are dynamic projects and have tight budgets. Another paper of Kautz, Hansen and Thaysen [5] has investigated the suitability for IDEAL model for small software enterprises, though it was actually based on the experiences of large organizations. IDEAL model was developed by Software Engineering Institute at Carnegie-Mellon University, USA (SEI). This model, as originally conceived, was a life-cycle model for software process improvement based upon the Capability Maturity Model (CMM) for software, and for this reason, this model uses process improvement terms. IDEAL model defines five phases: Initiating, Diagnosing,

Establishing, Acting and Learning. The Authors implemented this model in a small Danish company named 'NP' and deployment of IDEAL model took 10 weeks for them, 'one week initiating, three weeks diagnosing, and six weeks establishing and acting in parallel. McGuire & McKeown [6] showed how an ISO 9001 certified company adopted CMM level 2 first and later adopted level 3. The authors pointed that ISO 9001 had similarities with CMM level 2; however, differences were more pronounced and profound with CMM level 3. McGuire & McKeown placed a UK-based software company's example in this regard and examined that it takes five (5) steps to adopt CMM in an ISO environment, which are: 1) Establish a Software Engineering Process Group to change the culture of the organization, 2) Perform a gap analysis between ISO 9001 practices and CMM key practices, 3) Make a plan, schedule and detail the specific tasks, target a timeframe, 4) Provide training that address the CMM specific roles, and finally 5) Measure the improvement.

Each of the companies is following their own software development policy and software development standard if any. Some of the companies got certified that they are following ISO9000-1/2/3: 1994/2000 for their software development. Some of are trying to upgrade their certification up-to CMM level. But there are no unique guidelines for software development for all companies as well as for the nation so that each of the software developer can follow the unique standard guidelines throughout the country. These rules of software development come through the software development life cycle (SDLC). To develop international standard software, it is important to know the SDLC.

**Software Development Life Cycle (SDLC)**

The process of software development is often modeled as a series of stages that define the software life cycle. Software Development Life Cycle (SDLC) is an approach to develop a software product that is characterized by a linear sequence of steps that progress from start to finish. The SDLC model is one of the oldest systems development models and is still probably the most commonly used. The six general steps are: 1) evaluate existing system/software; 2) define new system requirements; 3) design system; 4) develop new system; 5) implementation; and, 6) maintain. SDLC is explained in Figure 1.

Software development life cycle models include prototyping, evolutionary prototyping, incremental development, spiral model, and V model. The use of these models is for the most part confined to the overall management of the project. However, projects are now considered better controlled if the model best suited to them.

Some experienced and highly respected project leaders and programmers consider rigid application of lifecycle plans to be a theory that does not work well in practice. Linux Torvalds, the very highly regarded project leader of the Linux Kernel, made the following statement on the Linux kernel mailing list: "No major software project that has been successful in a general marketplace (as opposed to niches) has ever gone through those nice lifecycles they tell you about in Computer Science classes" [7].

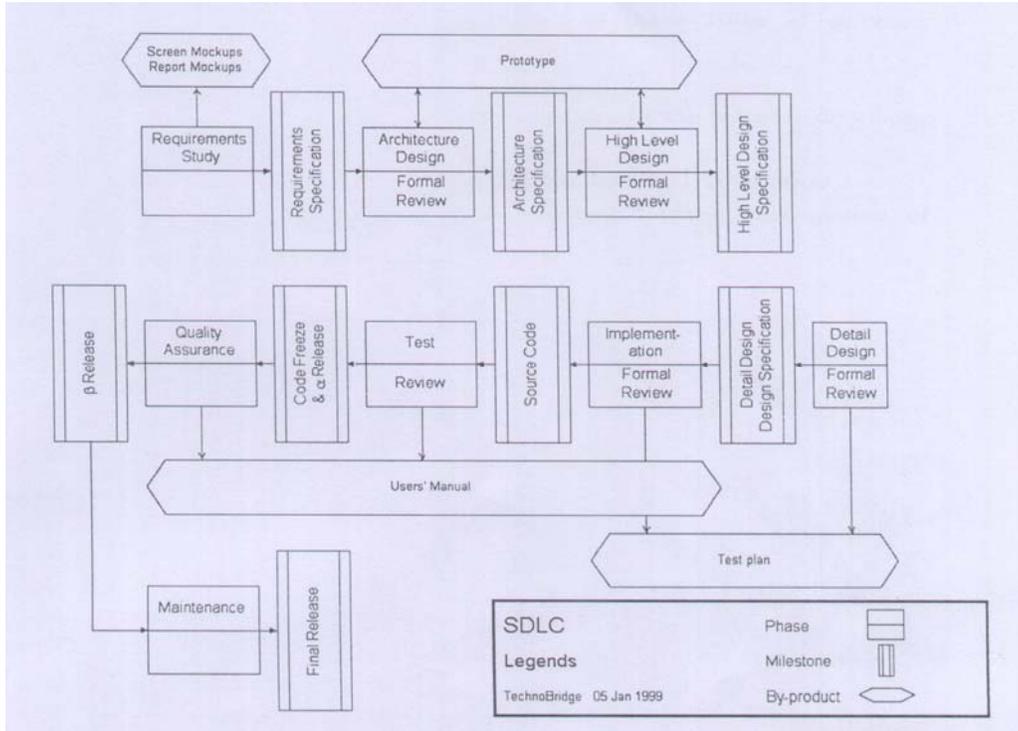

**Figure 1. Software Development Life Cycle**

## RESULT AND DISCUSSION

### Business Domain of the Surveyed Company

From our sample we observed that about 78.3% of the companies doing business on software development and sell and remaining 21.7% doing business on both software development and hardware sale. It is clear that our selected companies are doing software development and sale business. Figure 2 shows the nature of the business involved by the surveyed firms.

**Table 2. Descriptive Statistics on Surveyed Companies**

| | Descriptive Statistics | | |
|---|---|---|---|
| | | Mean (year) | Std. Deviation |
| 1. | How long company are doing business | 7.065 | 4.621 |
| 2. | How long an employee service with a firm | 5.022 | 6.380 |
| 3. | Number of people working for company | 25.289 | 17.991 |
| 4. | Total software developed | 12.030 | 18.691 |

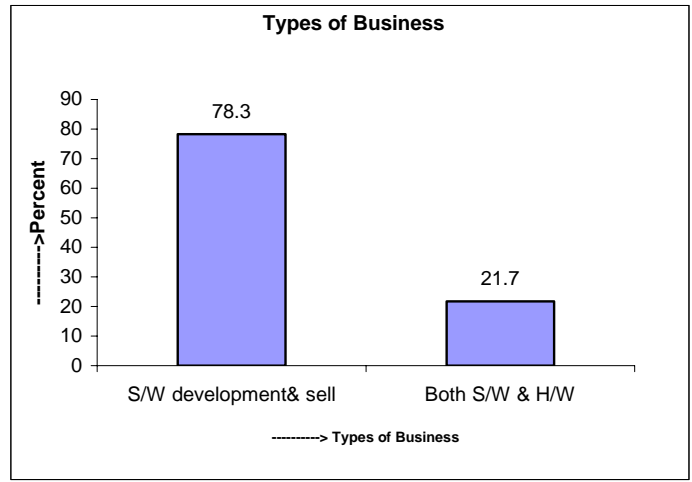

**Figure 2. Types of Business Doing the Selected Firm**

Table 2 shows the average years of business of the surveyed firms as 7.07 years. Their minimum and maximum years of business are 1 to 18 years respectively. Most of the firms are doing business for 1 to 3 years whereas few companies are doing business for 18 years. So they are not doing business for a long time. Similarly, average employees of these firms are 25 persons. Minimum and maximum employees are 5 and 55 respectively and average year of doing jobs in a company is 5 which indicate that most of them are small scale software firms. These firms on the average developed 12 (twelve) software during their existing period. They developed software ranging 2 to 20 in their life.

**Standardize Quality Assurance Certification**

To export software in international marker it is necessary to show that software are developed following a particular standard or set of rules like ISO 9001-3 or CMM. The software Engineering Institute's Capability Maturity Model (CMM) provides a well known benchmark of software process maturity. The CMM is a good perspective from which to asses the process framework. Having a matured process in place is far more important than merely passing an audit. A mature process would pass a surprise audit. The CMM defines five levels of software process maturity. It is reported that not a single company has achieved SEI's (Software Engineering Institute at Carnegie-Mellon University, USA) SW-CMM or CMMI level 3, though some of them exercise Personal Software Process to improve the quality of their development process [8]. CMM/CMMI level three is generally considered the minimum requirement for a company to be eligible to participate in global software industry.

Figure 3 reveals that about 44.4% companies got ISO 9001 certification for their software design, development, and maintenance services. About 11.1% companies got ISO 9002 certification, 5.6% companies got ISO 9003 certification, 8.3% companies got others (CMM) certification and 30.6% companies have no certification. We have to explore the standard followed by these 30.6% companies. We don't know exactly what they are following for their software development.

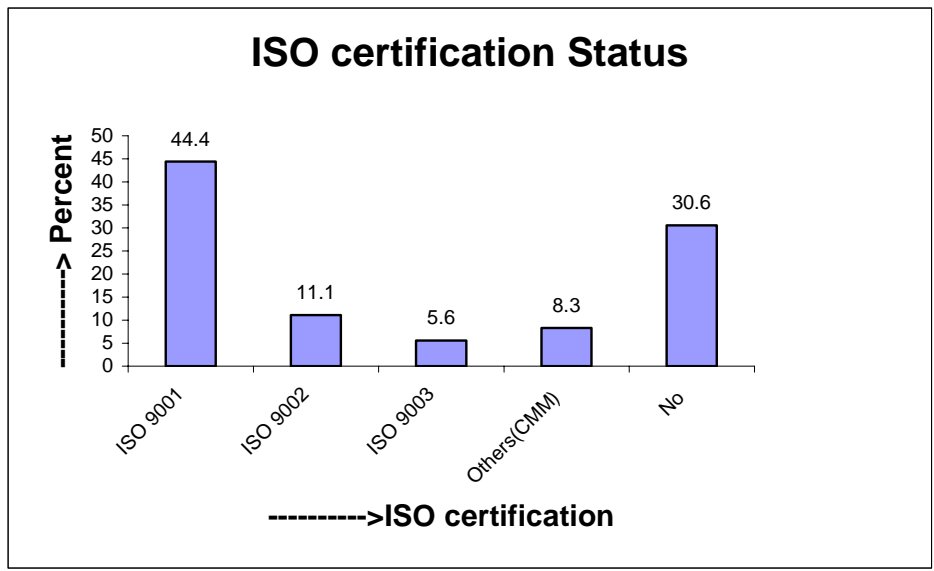

**Figure 3. ISO Certification Status of the Companies**

**Software Quality Control**

**Table 3. Percentages of Distributions of Software Quality Control Activities**

| SL. No. | Various stage of software development (N=50) | Yes (%) | No (%) |
|---|---|---|---|
| 1. | Have any quality manual for software development? | 83.7 | 16.3 |
| 2. | Do you have any quality policy for software development as a company? | 97.8 | 2.2 |
| 3. | Does the project follow a documented procedure to control the change of configuration items? | 78.6 | 21.4 |
| 4. | Have any written guideline for traceout the different software product? | 55 | 45 |
| 5. | Does the project follow a written organizational policy for implementing any software quality assurance (SQA) activities? | 52.5 | 47.5 |
| 6. | Have any documented procedure for process improvement? | 54.8 | 45.2 |

For the measurement of software quality control, we have some quarries to the software companies. The result of the quarries is shown in Table 3 which reveals that about 83.7% companies have quality manual. They strictly follow the quality manual during development of their software and 16.3% company response that have no quality manual. About 97.8% companies argue that they have a quality policy for software development as a company and 2.2% answered that they have no quality policy. 78.6% companies' project follows a documented procedure to control the change of configuration items and 21.4% does not follow the documented procedure. 55% companies said that they have a written guideline for trace out the different software product and 45% said they have not. Software quality assurance is a very important factor in software development. In our case study 52.5% companies follow a written organizational policy for implementing any software quality assurance (SQA) activities and 47.5% does not follow the policy. It is clear form Table 2 that about 54.8% companies have documented procedure for process improvement and 45.2% companies have no documented procedure for process improvement.

Table 4. Percentages of Distribution of Software Testing Activities

| Sl. No | Various stage of software development(N=50) | Yes (%) | No (%) |
|---|---|---|---|
| 1. | Have any test plan for the testing of the software is being developed? | 82.9 | 17.1 |
| 2. | Does the test plan for the testing of the software is reviewed by the technical manager/project manager? | 85.4 | 14.6 |

Table 4 indicates that 82.9% companies have a test plan for the testing of the software is being developed and 17.1% companies have not. It is also observed that about 85.4% companies reviewed their software for testing by the technical manager/project manager and 14.6% does not test any more.

**Software Development Process**

Table 5. Percentages of Distributions of Software Development Process Activities

| SL. No | Various stage of software development(N=50) | Yes (%) | No (%) |
|---|---|---|---|
| 1. | Have any written procedure/guideline for software design? | 75.6 | 24.4 |
| 2. | Have any written procedure/guideline for code generation? | 66.7 | 33.3 |

From Table 5 it is clear that 75.6% companies have a written procedure/guideline for software design and 24.4% have no guideline. About 66.7% companies have a written procedure/guideline for code generation and 33.3% have no guideline.

From Table 6 it is clear that though some companies have no ISO certification but 91.9% companies have a project planning procedures and 8.1% companies have no planning procedure which is very much insignificant.80.5% companies have a documented estimate for use in planning and tracking the software project. About 70% companies follow a written organizational policy for planning of the software project. 63.2% companies have a configuration management plans.

**Table 6. Percentages of Distributions of Project Planning Activities**

| SL. No. | Various stage of software development(N=50) | Yes (%) | No (%) |
|---|---|---|---|
| 1. | Have any Project Planning Procedures? | 91.9 | 8.1 |
| 2. | Do the estimates (size, cost, and schedule) are documented for use in planning and tracking the software project? | 80.5 | 19.5 |
| 3. | Does the project follow a written organizational policy for planning of the software project? | 70 | 30 |
| 4. | Have any configuration Management Plans? Or how can you address the Configuration Management? | 63.2 | 36.8 |

Software requirement Specification (SRS) is the most important key steps in designing a project. According to our data analysis, Table 7 shows that 90.7% companies have a guideline/procedure for requirement analysis/requirement specifications which his very positive and among them 70.7% companies has follow a written organizational policy for managing the system requirements analysis.

**Table 7. Percentages of Distributions of SRS Activities**

| SL. No | Various stage of software development(N=50) | Yes (%) | No (%) |
|---|---|---|---|
| 1. | Have any guideline/procedure for requirement analysis/requirement specifications? | 90.7 | 9.3 |
| 2. | If yes, does the project follow a written organizational policy for managing the system requirements analysis? | 70.7 | 29.3 |

**Business Expertise of the software Industry**

Response provided in Table 8 reveals the information that although most of the software company follows documented contract review procedure, they generally do not use any documented procedure for involving sub contractor. Even though few companies engaged sub contractor for developing their product they do not generally do not have any organizational policy to manage sub contractor. After sale services is very crucial for software business. But substantial percentage amount of software firm do not maintain documented procedure for after sales service. Most of the company gives priority on training which will be beneficial for the software industry in the long run. The statistical techniques to control and verify process capability of the organization need to introduce more in the software company in Bangladesh.

**Table 8. Percentages of Distributions of Business Expertise Factors**

| SL. No | Various stage of software development(N=50) | Yes (%) | No (%) |
|---|---|---|---|
| 1. | Have any documented/written contract review procedure (s) (before submission of a tender/the acceptance of a contract for software development)? | 69 | 31 |
| 2. | Have any documented procedure for amendment of contract review (if required to amendment the contract review)? | 50 | 50 |
| 3. | Have any documented procedure used to selecting a subcontractor for software? | 17.5 | 82.5 |
| 4. | Does the project follow a written organizational policy for managing software subcontractors? | 22 | 78 |
| 5. | Are the software subcontract activities reviewed by the project manager periodically? | 30 | 70 |
| 6. | Have any documented procedure for servicing the supplied products (s) after sales service? | 60.5 | 39.5 |
| 7. | Have any documented procedure for statistical technique used to controlling, and verify the process capabilities and product characteristics? | 64.1 | 35.9 |
| 8. | Have any team(s) is assigned to supervise and audit the above mentioned activities of the company? | 81.6 | 18.4 |

**Industry Growth and Manpower Scenario of Software Industry**

Now a days Software customers are clearly going global and are demanding quality. It is important for software organizations to understand all the rules for self-improvement and for doing business in the international marketplace. At present, more than fifty (50) software and IT service companies have been exporting their services to 30 countries in the world including USA, Canada, Middle East, Japan, Australia, South Africa and some of the South East Asian and European countries [9].

**Table 9. Growth of Software Export in Bangladesh**

| | 2000-2001 | 2001-2002 | 2002-2003 | 2003-2004 | 2004-2005 (till February, 2005 first eight months) |
|---|---|---|---|---|---|
| Export in US$ (in Million) | 2.24 | 2.8 | 4.2 | 7.2 | 7.38 |
| **Yearly Growth** | | 25% | 51% | 71% | 77% (over same period last year) |

Another important thing is that a knowledge based software industries required the qualified human resources in the market. The following table shows the academic background of the technical professionals employed software firms.

From the table 10, it is clear that only 44% professional are from Computer science graduate and 56% from other education background.

Table 10. Percentage of Total Technical Staffs in the Surveyed Software Firms

| Graduate in Non-IT subjects | 19% |
|---|---|
| Masters in Non-IT subjects | 23% |
| Computer Science/Engineering Graduate (3/4 years) | 35% |
| Master in Computer Science/Engineering | 9% |
| Diploma/Certificates courses in IT | 12% |
| Other | 2% |
| **Total** | **100%** |

Information provided in Table 11 proves that most of the software companies are aware of the training needs of their employee.

Table 11. Percentages on Distribution of Activities to Develop Skilled Manpower

| SL. No. | Various stage of software development(N=50) | Yes (%) | No (%) |
|---|---|---|---|
| 1. | Is the person in the project who is responsible for managing the requirement analysis activities trained in the procedure for managing requirement analysis? | 92.3 | 7.7 |
| 2. | Does the project manager review the requirement analysis output or product to be developed as per requirement? | 95 | 5 |
| 3. | Does the project manager review the activities for planning the software project? | 92.7 | 7.3 |
| 4. | Are the project personnel trained to perform the software configuration management activities for which they are responsible? | 73.2 | 26.8 |
| 5. | Does your organizational policy to meet the personnel training needs? | 75.6 | 24.4 |
| 6. | Does the software engineers/programmers trained before developing code (respective language)? | 85.7 | 14.3 |
| 7. | Are the training activities planned for the team members? | 90.2 | 9.8 |
| 8. | Do members of the software engineering group and other software related groups receive the training necessary to perform their roles? | 90.2 | 9.8 |
| 9. | Are training program activities reviewed with senior management on a periodic basis? | 80.5 | 19.5 |
| 10. | Are any measurements used to determine the quality of the training program? | 65.9 | 34.1 |

**CONCLUSION AND RECOMMENDATION**

Bangladesh poses some advantages regarding software development, which is very important to compete in global arena. One of them is low labor cost. Tjia [10] reports that cost per programmer is much lower in Bangladesh, which is more than 50% including overhead compared to India. High programmer productivity and wide spread of English have significant influence in software industry of Bangladesh. According to our survey 54% of the workforces are graduates from non-IT subjects and diploma/certificate courses in IT. Generally, they are trained in professional IT courses from different internationally recognized IT institutions established in Bangladesh. This justifies the analytical and technological ability of Bangladeshi people. With this competency, Bangladeshi software companies have to exploit the advantages of having no major cultural differences with other client countries as have reflected so far.

Software process and process improvement is the key to success for a software company in software business. Without a process, a software organization is not capable of producing high quality software. Without defined and measured software process, a software company fails to determine its business goal. Software companies of Bangladesh are facing the problem of adaptation of specific software process model. Our research has identified some problems. These are lack in target set for software process and improvement, lack in involvement of quality control activities, and lack of standard business expertise practice.

This paper is a quantitative study and is based on questionnaires and interviews of software companies' personnel. The total amount of software companies in Bangladesh is more than 250 if we include the software company those are not enlisted with BASIS and we have collect data from around 50 companies. Our result does not show the whole picture of Bangladesh. Therefore, the study is exploratory. The results are not conclusive at this stage. A further study is needed to explore the extensive scenario of the software development standard of Bangladesh.